\documentclass[a4paper,fleqn]{cas-dc}

\usepackage[numbers,sort&compress]{natbib}

\def\tsc#1{\csdef{#1}{\textsc{\lowercase{#1}}\xspace}}
\tsc{WGM}
\tsc{QE}
\tsc{EP}
\tsc{PMS}
\tsc{BEC}
\tsc{DE}

\begin{document}
\let\WriteBookmarks\relax
\def\floatpagepagefraction{1}
\def\textpagefraction{.001}
\shorttitle{}
\shortauthors{Shubham et~al.}
\title [mode = title]{3D-Ising-type Magnetic Interactions Stabilized by the Extremely Large Uniaxial Magnetocrystalline Anisotropy in Layered Ferromagnetic Cr$_2$Te$_3$}
\author{Shubham Purwar}
\credit{Conceptualization, Methodology, Validation, Formal analysis, Investigation, Writing original draft, Writing review \& editing}
\author{Tushar Kanti Bhowmik}
\credit{DFT calculations, Methodology Validation, Writing review \& editi ng}
\author{Soumya Ghorai}
\credit{Writing review \& editing}
\author{Setti Thirupathaiah}[orcid=0000-0003-1258-0981]
\ead{setti@bose.res.in}
\ead[URL]{www.qmat.in}
\credit{Conceptualization, Validation, Resources, Supervision, Project administration, Funding acquisition, Writing review \&
editing}

\affiliation{Department of Condensed Matter and Materials Physics, S. N. Bose National Centre for Basic Sciences, Kolkata, West Bengal 700106, India.}

\begin{abstract}
We investigate the magnetocrystalline anisotropy, critical behavior, and magnetocaloric effect in ferromagnetic-layered Cr$_2$Te$_3$. We have studied the critical behavior around the Curie temperature ($T_C$) using various techniques, including the modified Arrott plot (MAP), the Kouvel-Fisher method (KF), and critical isothermal analysis (CI). The derived critical exponents $\beta$ = 0.353(4) and $\gamma$ = 1.213(5) fall in between the three-dimensional (3D) Ising and 3D Heisenberg type models, suggesting complex magnetic interactions by not falling into any single universality class. On the other hand, the renormalization group theory, employing the experimentally obtained critical exponents, suggests 3D-Ising-type magnetic interactions decaying with distance as $J(r) = r^{-4.89}$. We also observe an extremely large uniaxial magnetocrystalline anisotropy energy (MAE) of $K_u=2065$ kJ/m$^3$, the highest ever found in any Cr$_x$Te$_y$ based systems, originating from the noncollinear ferromagnetic ground state as predicted from the first-principles calculations.  The self-consistent renormalization theory (SCR) suggests Cr$_2$Te$_3$ to be an out-of-plane itinerant ferromagnet. Further, a maximum entropy change of -$\Delta S_{M}^{max}\approx$ 2.08 $J/kg-K$ is estimated around $T_C$ for the fields applied parallel to the $c$-axis.
 \end{abstract}


\maketitle

\section{\label{sec:level1}Introduction}

The discovery of intrinsic long-range ferromagnetism in two-dimensional (2D) layered materials has ignited considerable research interest due to their potential technological applications in low-power spintronic devices~\cite{Heron2011,Geim2013,Gong2017,Fei2018,Wang2019,Gibertini2019}. On the other hand, the Mermin-Wagner theorem proposes certain limitations to have long-range ferromagnetic interactions in the 2D magnets due to thermal fluctuations at finite temperatures~\cite{Mermin1966}. In this regard, the magnetocrystalline anisotropy energy (MAE) becomes crucial to overcome the thermal fluctuations and for the realization of 2D long-range ferromagnetism in layered materials. Recent investigations have focused on enhancing the long-range magnetic ordering in several 2D layered magnets such as CrI$_3$~\cite{Huang2017}, Cr$_2$Ge$_2$Te$_6$~\cite{Liu2017}, and Cr$_2$Si$_2$Te$_6$~\cite{Williams2015}. Nevertheless, the peculiar magnetic behavior observed in chromium-based tellurides (Cr$_x$Te$_y$) have drawn significant research attention~\cite{Zhang2020,Wang2022,Liu2019}.

Under different synthesis conditions and chromium concentrations, Cr$_{x}$Te$_y$ exhibits diverse crystal structures owing to the Cr vacancies. For instance, CrTe, Cr$_5$Te$_6$, and  Cr$_{7}$Te$_8$ exhibit hexagonal crystal structure~\cite{Hashimoto1969, Zhang2022, Liu2023}, Cr$_{3}$Te$_4$ and Cr$_{5}$Te$_8$ show monoclinic phase~\cite{Yamaguchi1972,Chattopadhyay1994,Purwar2023},  and Cr$_{2}$Te$_3$ and CrTe$_2$ show trigonal crystal structure~\cite{Freitas2015, Bian2021}. Thus, the crystal structure of Cr$_x$Te$_y$ is highly sensitive to the Cr concentration present in the system, $\it{viz.}$  Cr$_5$Te$_8$ shows monoclinic phase for 38.5 - 40.4 \% of Cr, while it shows trigonal structure for 38 - 37.5 \% of Cr~\cite{Ipser1983}. Moreover, the Cr concentration significantly affects the magnetic properties as well.  Though most of the Cr$_x$Te$_y$ systems show ferromagnetic (FM) ordering, CrTe$_3$ (Cr$_{0.67}$Te$_2$) exhibits an antiferromagnetic ordering~\cite{McGuire2017}, their Curie temperatures are highly sensitive to the Cr concentration. For instance, Cr$_5$Te$_8$ (Cr$_{1.25}$Te$_2$) shows FM order below $T_C\approx230$ K~\cite{Liu2019, Mondal2019},  Cr$_{3}$Te$_4$ (Cr$_{1.5}$Te$_2$) shows FM order below $T_C\approx316$ K~\cite{Yamaguchi1972}, Cr$_{7}$Te$_8$ (Cr$_{1.75}$Te$_2$) shows FM order below $T_C\approx361$ K~\cite{Hashimoto1969}, CrTe$_2$ shows FM order below $T_C\approx310$ K~\cite{Freitas2015}, and Cr$_{2}$Te$_3$ (Cr$_{1.33}$Te$_2$) shows FM order below $T_C\approx180$ K~\cite{Bian2021}. Not only the Curie temperature, the magnetic exchange interactions below the Curie temperature are also sensitive to the Cr concentrations. For instance, below $T_C$, Cr$_5$Te$_8$ shows 3D-Ising type magnetic interactions, Cr$_4$Te$_5$ shows 3D-Heisenberg type magnetic interactions~\cite{Zhang2020}, and CrTe shows quasi-2D-Heisenberg type magnetic interactions~\cite{Liu2023}.

 \begin{figure*}[ht]
    \centering
    \includegraphics[width=\linewidth]{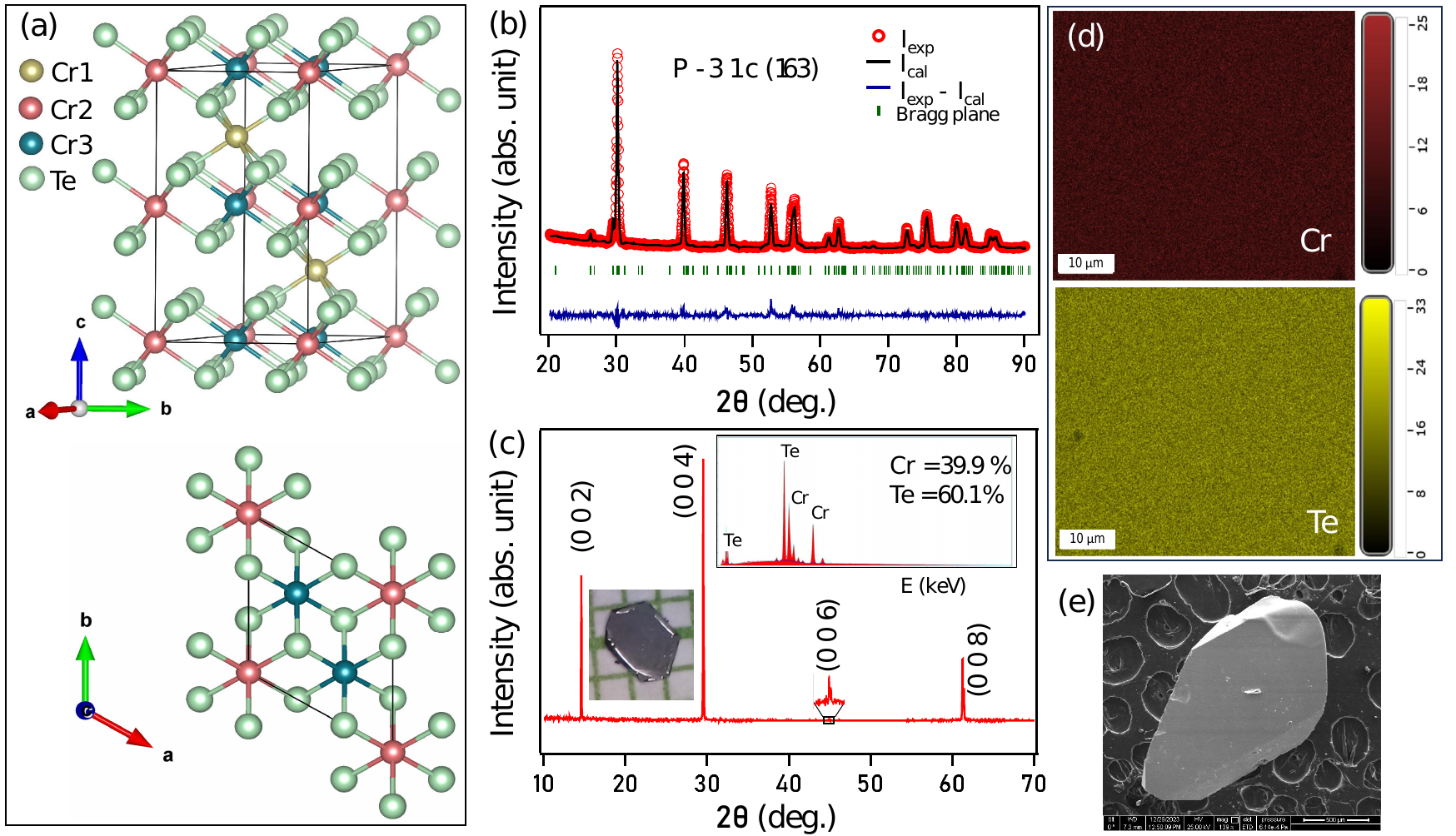}
    \caption{(a) Schematic representations of the Cr$_2$Te$_3$ crystal structure, obtained from the Rietveld refinement. (b) XRD pattern of Cr$_2$Te$_3$ crushed crystals measured at 300 K overlapped with the Rietveld refinement. (c) XRD pattern of Cr$_2$Te$_3$ single crystal. Insets in (c) show EDXS data demonstrating the actual chemical composition of as-prepared crystals and the photographic image of a typical Cr$_2$Te$_3$ single crystal. (d) Elemental mapping of Cr$_2$Te$_3$ for Cr and Te using EDXS. (e) Scanning electron microscopic (SEM) image of Cr$_2$Te$_3$ single crystal.}
    \label{1}
\end{figure*}

Although the ground state magnetism of Cr$_2$Te$_3$ single crystals has been discussed earlier~\cite{Jiang2022}, no study is available on Cr$_2$Te$_3$ discussing the magnetic exchange interactions across the Curie temperature. Furthermore, Cr$_2$Te$_3$ thin films have exhibited an anomalous Hall effect~\cite{Wen2020,Luo2022}, while the topological Hall effect has been observed in hetero-junction devices~\cite{Jeon2022}. Thus, a better understanding of the magnetic interactions in Cr$_2$Te$_3$ is crucial to unravel the potential technological applications of these systems.

In this contribution, we discuss the magnetocrystalline anisotropy, critical behavior, and magnetocaloric effect in Cr$_2$Te$_3$. The critical behavior around the Curie temperature ($T_C$) has been studied using various techniques such as the modified Arrott plot (MAP), the Kouvel-Fisher method (KF), and critical isothermal analysis (CI). Interestingly, the derived critical exponents $\beta$ = 0.353(4) and $\gamma$ = 1.213(5) suggest complex magnetic interactions falling in between  the 3D Ising and 3D Heisenberg models across the $T_C$. On the other hand, the renormalization group theory suggests 3D-Ising type magnetic interactions stabilized by the extremely large magnetocrystalline anisotropy. We performed first-principles calculations to establish the origin of extremely large out-of-plane magnetic anisotropy. Our calculations predict a noncollinear ferromagnetic phase with a canted spin structure that results from the magnetic frustration~\cite{Abuawwad2022}, in the ground state with dominant ferromagnetic ordering along the $c$-axis, while an antiferromagnetic ordering is found in the $ab$-plane, leading to extremely large magnetocrystalline anisotropy.  Further, a maximum entropy change of -$\Delta S_{M}^{max}\approx$ 2.08 $J/kg-K$ is estimated around $T_C$ for the fields applied parallel to the $c$-axis.


\section{Experimental and First-principles Calculation Details}

High-quality single crystals of Cr$_2$Te$_3$ were grown using the chemical vapor transport (CVT) technique by taking $3 : 4$ molar ratio of Chromium (Cr, 99.99\%, Alfa Aesar) and Tellurium (Te, 99.999\%, Alfa Aesar) powders and employing iodine as the transport agent. The growth process  involved a three-week thermal heat treatment in a two-zone horizontal tube furnace, with the source-zone maintained at 1000 $^{\circ}$C and the growth-zone kept at 820 $^{\circ}$C~\cite{Hashimoto1971}. The as-grown single crystals, sized 3$\times$2 mm$^2$, were looking shiny and flat [see the inset of Fig.~\ref{1}(c)]. Surface morphology and elemental compositions were studied by using the scanning electron microscope (SEM) and energy dispersive x-ray spectroscopy (EDXS). We find that the as-prepared single crystals have a Cr to Te ratio of 0.663(6):1, leading to the exact chemical formula of Cr$_{1.99}$Te$_3$. Powder x-ray diffraction (XRD) analysis was conducted using a Rigaku X-ray diffractometer (SmartLab, 9kW) with Cu K$_\alpha$ radiation (wavelength = 1.5406 \AA). Magnetic properties studies, $M(T)$ and $M(H)$, were performed using the physical property measurement system (PPMS, 9 Tesla DynaCool, Quantum Design). The magnetization isotherms $M(H)$ were systematically measured with 2 K interval around the Curie temperature to facilitate the critical behavior analysis.

Density functional theory (DFT) calculations on Cr$_2$Te$_3$ were performed using the generalized gradient approximation (GGA) of Perdew,
Burke and Ernzerhof (PBE) exchange and correlation functionals~\cite{Perdew1996} as implemented in the Quantum Espresso (QE) simulation package~\cite{Giannozzi2009}. Brillouin zone sampling was done over a 12$\times$12$\times$6 Monkhorst-Pack $k$-grid. The electronic wave function is expanded using the plane waves up to a cutoff energy of 40 Ry. To accurately determine the spin moment distribution, our approach involved incorporating scalar-relativistic and relativistic pseudopotentials to account the  within the self-consistent field (SCF) calculations. We explored both collinear and non-collinear spin structures, the latter involving optimization of the spin angle with respect to the energy.

\begin{figure}[t]
    \centering
    \includegraphics[width=0.95\linewidth]{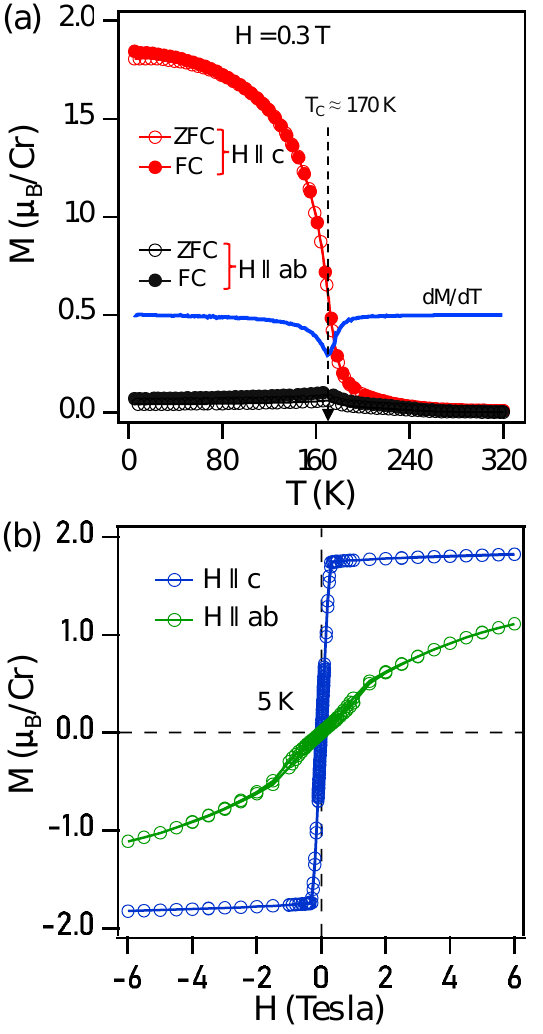}
    \caption{(a) Temperature-dependent magnetization $M(T)$ measured for $H\parallel c$ and $H\parallel ab$  in the ZFC and FC modes under an applied of $H=$0.3 T. In (a), the blue-curve displays first derivative of magnetization with respect to temperature, $\frac{dM}{dT}$, showing a dip at the $T_C\approx170$ K. (b) Magnetization isotherms [$M(H)$] measured at 5 K for $H\parallel c$ and $H\parallel ab$ orientations.}
    \label{2}
\end{figure}

\begin{figure}[t]
    \centering
    \includegraphics[width=\linewidth]{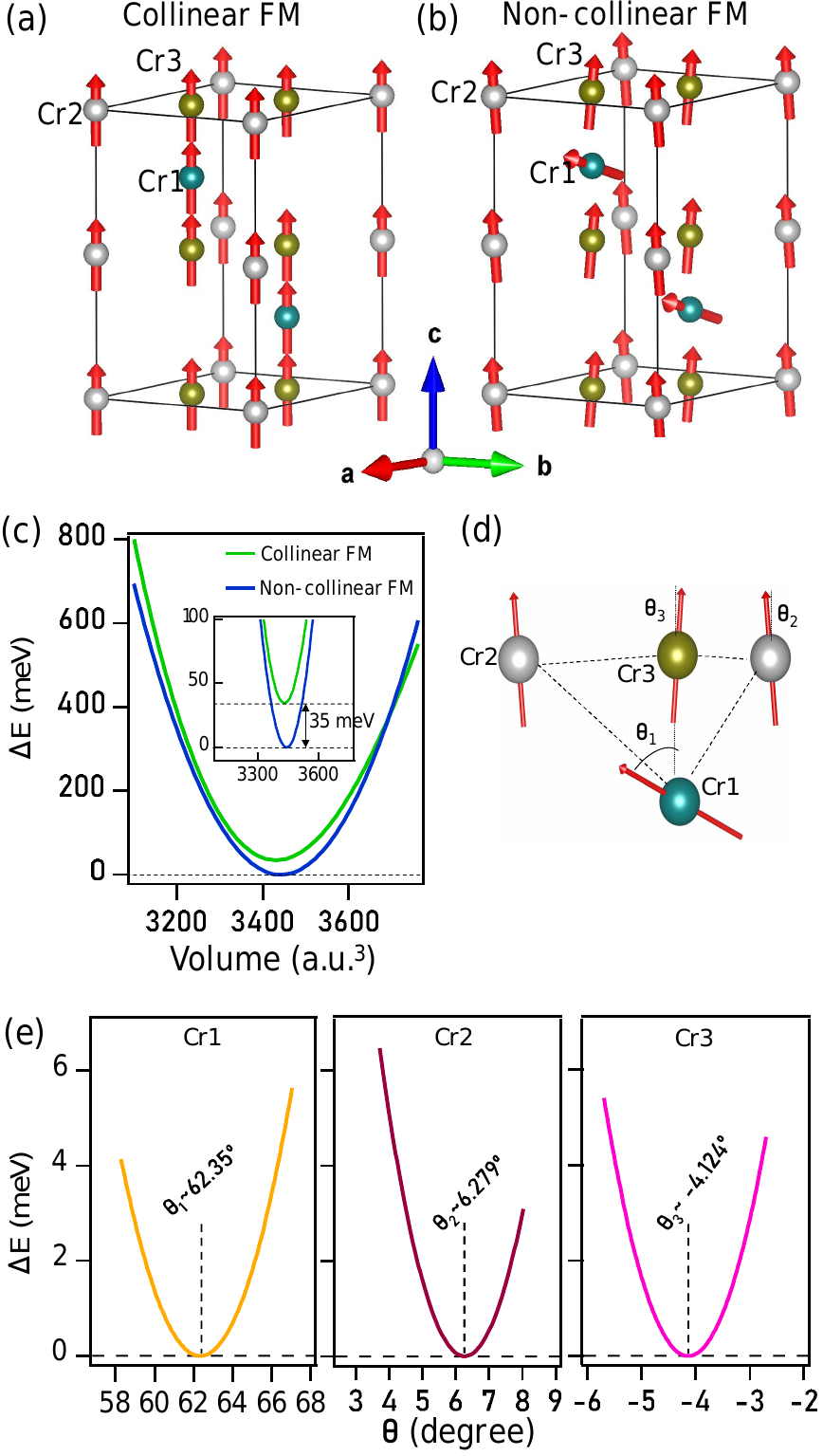}
    \caption{Schematic representations of the collinear ferromagnetic (FM) (a) and noncollinear ferromagnetic (b) configurations. (c) Relative total energy ($\Delta$E),  with respect to the ground state energy,  plotted as a function of unit cell volume for both magnetic configurations, obtained using the DFT calculations. (d) Schematic diagram of the canting spin angles of Cr(1), Cr(2), and Cr(3) atoms with respect to the $c$-axis. (e) Relative total energy ($\Delta$E), with respect to the ground state energy,   plotted as a function of spin angles for Cr(1), Cr(2) and Cr(3) atoms.}
    \label{2a}
\end{figure}

\section{Results and Discussion}

\subsection{Structural analysis}
The unit cell of Cr$_2$Te$_3$ is schematically shown in the top-panel of Fig.~\ref{1}(a), in which distinct chromium atoms are labeled by Cr(1), Cr(2), and Cr(3). The intercalated Cr(1) atoms reside within the van der Waals gap created by two CrTe$_2$ layers, exhibiting ordered vacancies. Cr(2) and Cr(3) are within the CrTe$_2$ layers, but only Cr(3) has the nearest neighboring Cr(1) atoms along the $c$-axis. On the other hand, Cr(1) has two Cr(3) nearest neighbours along the $c$-axis but has no nearest neighbors in the $ab$-plane. The bottom-panel of Fig.~\ref{1}(a) depicts projected crystal structure of  Cr$_2$Te$_3$ onto the $ab$-plane, showing the intertwined honeycomb lattice with Cr(2), Cr(3), and Te atoms.   Fig.~\ref{1}(b) displays the X-ray diffraction (XRD) pattern of the crushed single crystals measured at room temperature. The XRD data confirms the trigonal crystal structure of Cr$_2$Te$_3$ having the space group of $P\bar{3}1c$ (163). The derived lattice parameters using the Rietveld refinement,  $a$ = $b$ = 6.7989(7) \AA, $c$ = 12.1074(7) \AA, and the angles $\alpha = \beta$ = 90$^{\circ}$, $\gamma$ = 120$^{\circ}$,  are close to the previous reports~\cite{Hamasaki1975,Ipser1983,Chattopadhyay1994}. Fig.~\ref{1}(c) displays the XRD pattern taken on the Cr$_2$Te$_3$ single crystal,  indicating that the crystal growth is parallel to the $(00l)$ Bragg's plane. Using the Rietveld refinement, bond lengths of Cr$_1$-Cr$_3$, Cr$_2$-Cr$_3$, Cr$_2$-Cr$_2$, and Cr$_3$-Cr$_3$ are determined as 3.0269 \AA, 3.9254 \AA, 6.7990 \AA, and 3.9254 \AA, respectively. Additionally, the bond angles of Cr$_2$-Te$_1$-Cr$_3$, Cr$_1$-Te$_1$-Cr$_2$, and Cr$_1$-Te$_1$-Cr$_3$ are found as 91.213$^{\circ}$, 129.596$^{\circ}$, and 67.542$^{\circ}$, respectively. Fig.~\ref{1}(d)  exhibits the elemental mapping of Cr and Te, confirming the uniform chemical composition of the studied crystal within the measured surface range of 50$\times$45 $(\mu m)^2$. Fig.~\ref{1}(e) shows scanning electron microscopy (SEM) image of Cr$_2$Te$_3$, displaying a very flat surface morphology. Thus, the EDXS and SEM data confirm the homogeneity and high-quality of the studied single crystals.

\begin{figure}[t]
    \centering
    \includegraphics[width=\linewidth]{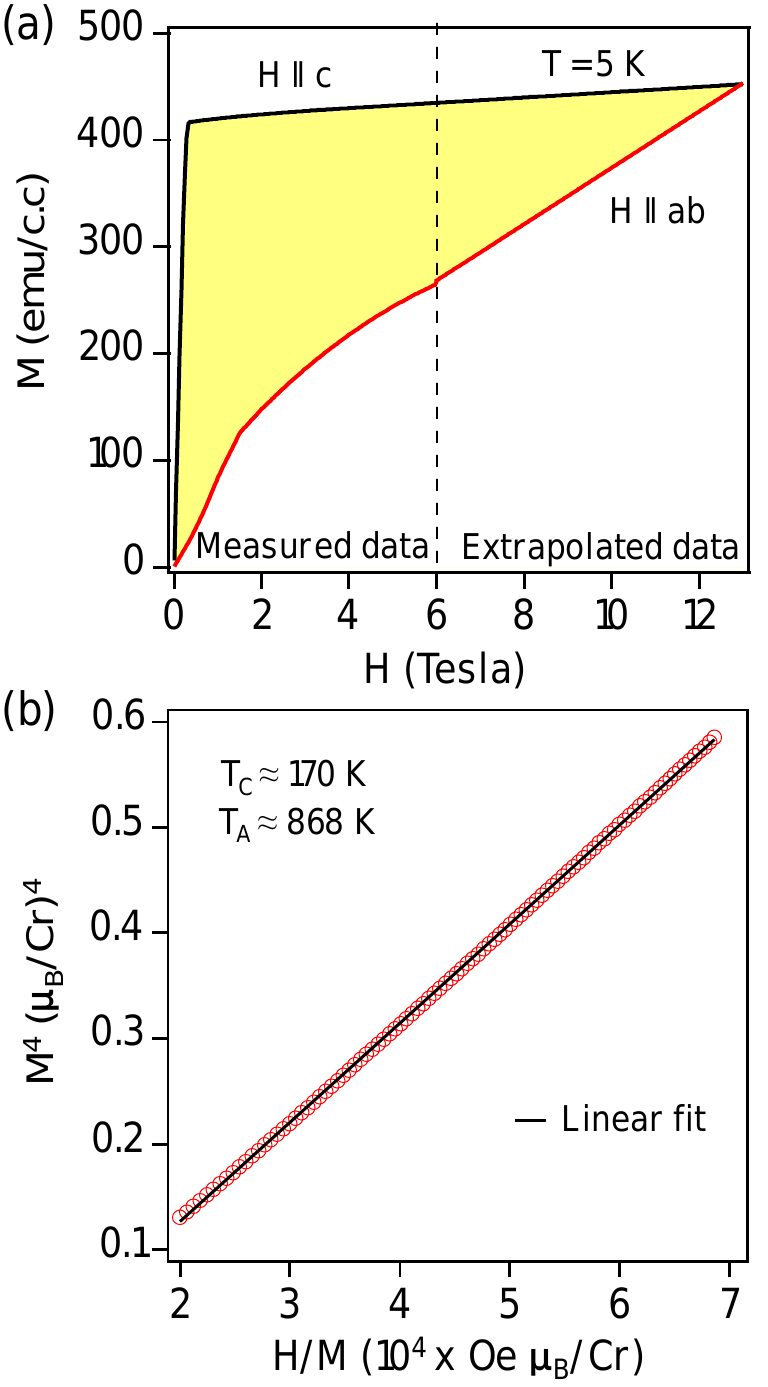}
    \caption{(a) Magnetization isotherms [$M(H)$] measured at 5 K for $H\parallel c$ and $H\parallel ab$. The yellow shaded area represents the magnetocrystalline anisotropy energy (MAE).  (b) Plot of M$^4$ $vs.$ $H/M$}
    \label{SCR}
\end{figure}

\subsection{Magnetic Properties}

Temperature-dependent magnetization [$M(T)$] of Cr$_2$Te$_3$ measured at $H=$ 0.3 T for both the $H\parallel c$ and $H\parallel ab$ orientations  are shown in Fig.~\ref{2}(a) taken in field-cooled (FC) and zero-field-cooled (ZFC) modes. From Fig.~\ref{2}(a), we observe a ferromagnetic-like magnetic transition at around $T_C\approx$ 170 K. The same is confirmed from the overlapped $dM/dT$ data in which a dip is noticed at around 170 K. Most interestingly, we observe an out-of-plane ($H\parallel c$) saturation magnetization (1.8 $\mu_B/Cr$) that is almost 20 times higher than the in-plane ($H\parallel ab$) saturation magnetization (0.08 $\mu_B/Cr$), indicating an extremely large magnetic anisotropy in this system with an easy-axis of magnetization parallel to the $c$-axis. To further confirm the magnetic anisotropy, we measured magnetization as a function of the field [$M(H)$] for both orientations as shown in Fig.~\ref{2}(b) at 5 K of the sample temperature. From Fig.~\ref{2}(b), we clearly notice a spontaneous magnetization of 1.8 $\mu_B/Cr$ for $H\parallel c$ around the zero applied field with negligible coercivity, suggesting Cr$_2$Te$_3$ to be an out-of-plane soft-ferromagnet. On the other hand, for $H\parallel ab$, the lower magnetic fields sustain a canted FM state between neighboring Cr(2) and Cr(3) atoms. 

\begin{figure*}[t]
    \centering
    \includegraphics[width=\linewidth]{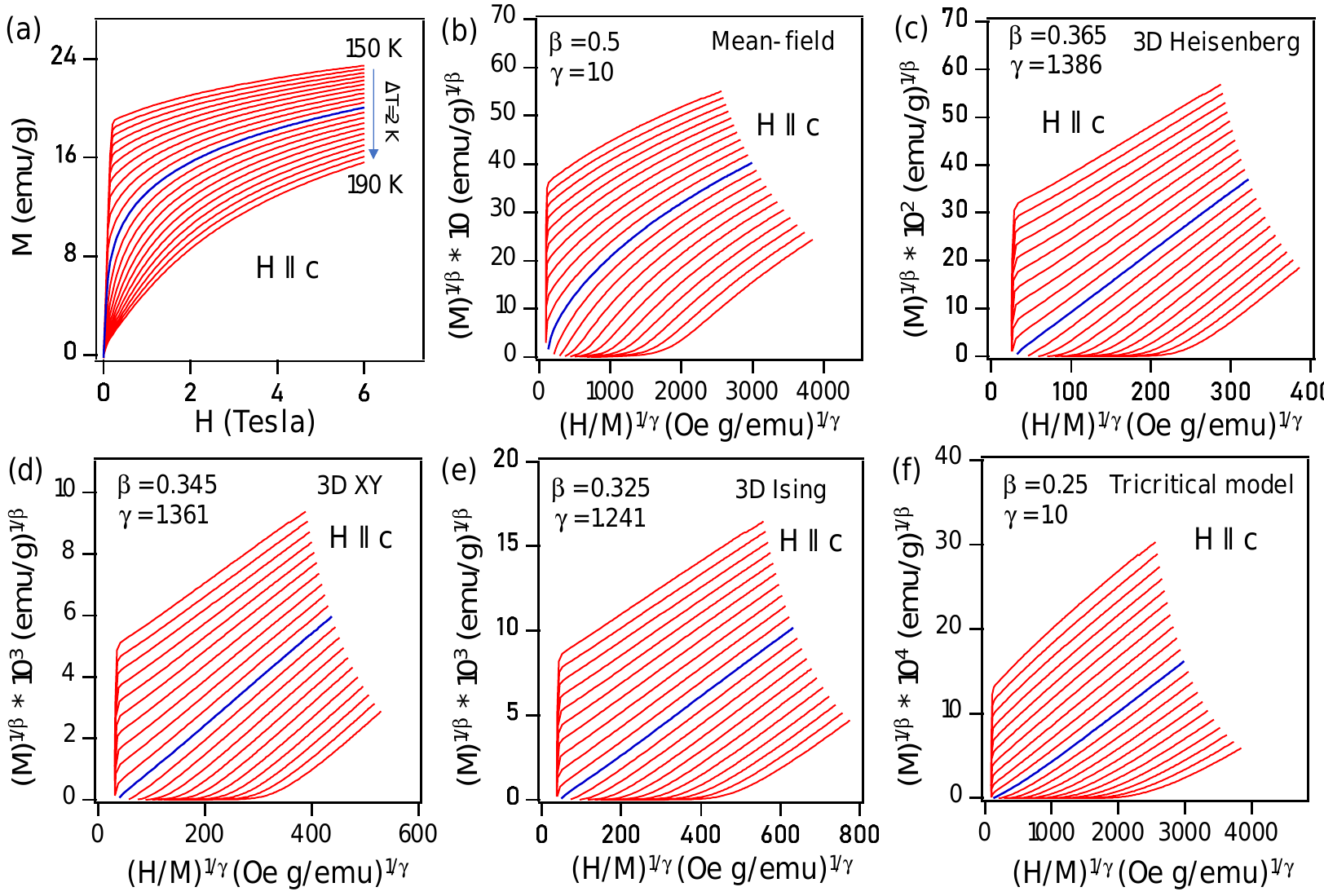}
    \caption{(a) Isothermal magnetization $M(H)$ measured between 150 and 190 K with an interval of 2 K for $H\parallel c$. (b) Arrot plots of M$^{2}$ $vs.$ $(H/M)$ for the Landau mean-field model. Modified Arrott plots of M$^{1/\beta}$ $vs.$ (H/M)$^{1/\gamma}$ for (c) 3D-Heisenberg model, (d) 3D-XY model, (e) 3D-Ising model, and (f) Tricritical mean-field model. In the figures, the blue-colored data is taken at the $T_C\approx$ 170 K.}
    \label{3}
\end{figure*}

\begin{figure}[t]
    \centering
    \includegraphics[width=\linewidth]{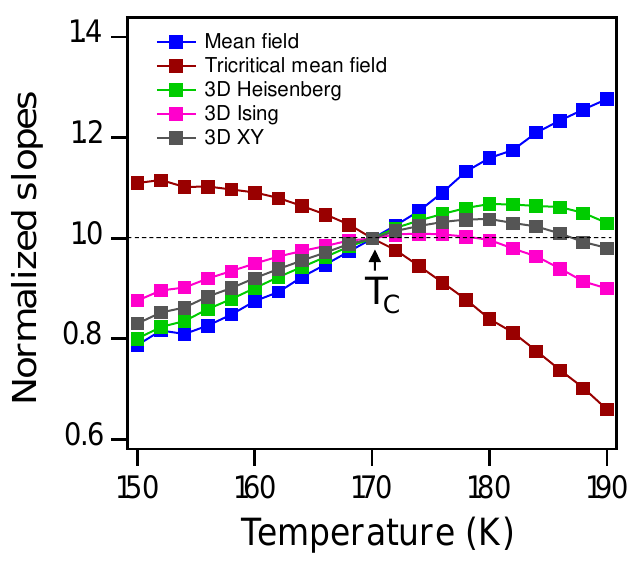}
    \caption{Temperature dependent normalized slopes [NS=S(T)/S(T$_C$)] plotted as a function of temperature, for various models, derived from the data shown in Figs.~\ref{3}(b) -~\ref{3}(f). See the text for more details.}
    \label{slope}
\end{figure}

\begin{figure*}[t]
    \centering
    \includegraphics[width=\linewidth]{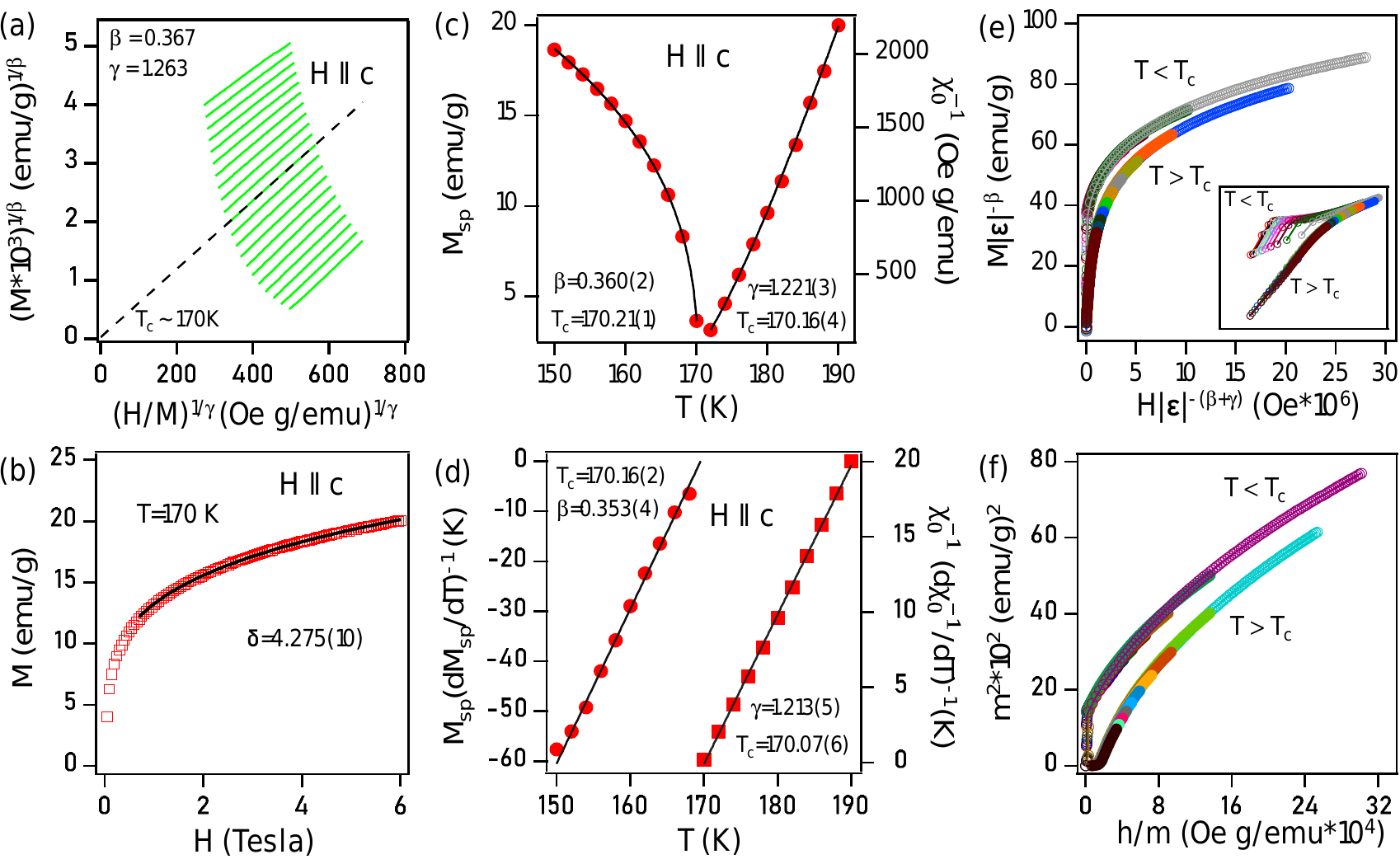}
    \caption{(a) Modified Arrott plot (MAP) of M$^{1/\beta}$ $vs.$ (H/M)$^{1/\gamma}$ at high-field regions  with critical exponents, $\beta$ = 0.367 and $\gamma$ = 1.263. (b) Isothermal magnetization $M(H)$  data taken at T$_C$ = 170 K (c) Temperature dependent spontaneous magnetization M$_{sp}$(T) (left axis) and inverse initial susceptibility $\chi_0^{-1}(T)$ (right axis). (d) Kouvel-Fisher plot: temperature dependence of $M_{sp}(T)[dM_{sp}(T)/dT]^{-1}$ (left axis) and $\chi_0^{-1}(T)[d \chi_0^{-1}(T)/dT]^{-1}$ (right axis). (e) Scaling plot of normalized magnetization vs. normalized field. (f) Renormalized magnetization ($m$) and field ($h$) are plotted in the form of $\it{m^2}$ $vs.$ $\it{h/m}$ below and above T$_C$.  Inset in (e) shows the same plot of (e) in log-log scale.}
    \label{4}
\end{figure*}

We performed density functional theory (DFT) calculations  to reveal the ground state magnetic structure of Cr$_2$Te$_3$. Initially we focused on examining the magnetic coupling among various types of Cr atoms, including the collinear ferromagnetic (FM) and non-collinear FM configurations as shown in Figs.~\ref{2a}(a) and  ~\ref{2a}(b), respectively. The DFT calculations are done by including the spin-orbit coupling (SOC) effect. Our calculations suggest that the noncollinear ferromagnetic configuration [see Fig.~\ref{2a}(c)] has a lower ground state energy by about 35 meV  compared to the collinear FM configuration, thus the former is a more stable magnetic ground state than the latter in Cr$_2$Te$_3$. Distinct spin canting angles were observed for Cr(1), Cr(2), and Cr(3) deviating from the $c$-axis as schematically shown in Fig.~\ref{2a}(d). This investigation suggests substantial difference in the canted spin angles, between the intercalated Cr(1) atoms and  Cr(2) or Cr(3) of CrTe$_2$ layer~\cite{Bian2021}. Further, Cr(1) and Cr(2) display spin canting in the second quadrant, while Cr(3) exhibits spin canting in the first quadrant, as depicted in Fig.~\ref{2a}(d). Fig.~\ref{2a}(e) depicts the plots of relative total energy ($\Delta$E) $vs.$ spin canting angle, suggesting 62.35$^{\circ}$, 6.279$^{\circ}$, and -4.124$^{\circ}$ are the canted-spin angles of Cr(1), Cr(2), and Cr(3) atoms, respectively in the magnetic ground state. Therefore, the predicted canted angles indicate a dominant ferromagnetic ordering for the spins of Cr(2) \& Cr(3) of CrTe$_2$ layer and a dominant antiferromagnetic ordering for the intercalated Cr(1) atoms.  As a result, the noncollinear FM ground state in Cr$_2$Te$_3$ is stabilized by the magnetic frustration due to a competition between the in-plane AFM and the out-of-plane FM orders. These observations are consistent with predictions on similar systems~\cite{ANDRESEN1970,Bian2021}.

Next, with the help of magnetization isotherms [$M(H)$] measured at 5 K for both $H\parallel c$ and $H\parallel ab$ [see Fig.~\ref{SCR}(a)], we qualitatively estimated the magnetocrystalline anisotropy  energy (MAE) $K_u$ in our studied Cr$_2$Te$_3$ sample by measuring the area  (Yellow-shaded region) between $H\parallel c$ and $H\parallel ab$ curves, using the relation~\cite{Purwar2023},

\begin{equation}\label{Ku}
K_u=\mu_0\int_{0}^{M_s}[H_{c}(M)-H_{ab}(M)]~dM \\
\end{equation}
Here, $\mu_0$ is the vacuum permeability, M$_s$ represents saturation magnetization.  H$_{c}$ and H$_{ab}$ represent magnetic field applied along the out-of-plane and in-plane directions, respectively.

The derived value of $K_u=2065$ kJ/m$^3$ in this method is found to be the highest ever known MAE from any Cr$_x$Te$_y$ type systems. Moreover, this value is much larger than the K$_u$ values reported on many other 2D magnetic systems such as Cr$_2$Si$_2$Te$_6$ / Cr$_2$Ge$_2$Te$_6$) ($\approx$ 65 kJ/m$^3$ / $\approx$20 kJ/m$^3$)~\cite{Liu2019a},  Fe$_4$GeTe$_2$ (250 kJ/m$^3$)~\cite{Mondal2021},  CrBr$_3$ / CrI$_3$ ($\approx$ 86 kJ/m$^3$ /$\approx$ 300 kJ/m$^3$)~\cite{Richter2018}, and Fe$_3$GeTe$_2$ (1460 kJ/m$^3$)~\cite{leon2016magnetic}.  Thus, for the first-time we report an extremely large MAE from our studied Cr$_{2}$Te$_3$. Note here that we estimated the MAE value by linearly extrapolating the high-field region of the $M(H)$ data for $H\parallel ab$ [see Fig.~\ref{SCR}(a)] to find the overlapping filed position (12.8 T) with $H\parallel c$. Therefore, the MAE value $K_u=2065$ kJ/m$^3$ is the minimum value that we could estimate in this method. The actual $K_u$ value would be higher than what we estimated, if one is able to perform $M(H)$ measurements at very high applied magnetic fields and find the overlapping field position directly from the measurements. Nevertheless, we could also estimate the magnetocrystalline anisotropy energy using the DFT calculations. In agreement with the experimental value, we find an extremely large MAE for the noncollinear FM state of 8230 kJ/m$^3$. Whereas for the collinear FM state we find realatively very low MAE of 32 kJ/m$^3$.



\subsection{Itinerant Ferromagnetism}

To identify the nature of ferromagnetism in Cr$_{2}$Te$_{3}$ we employed Takahashi's self-consistent renormalization (SCR) theory which takes into account the conservation of zero-point spin fluctuations and thermal spin fluctuations~\cite{Takahashi2013}. According to SCR theory, the magnetization $M$ and the magnetic field $H$ at T$_C$ should obey the below relation,
\begin{equation}
  M^4 = \frac{1}{4.671} \left[\frac{T_C^2}{T_A^3}\right] \left(\frac{H}{M}\right)
 \label{Eqn1}
 \end{equation}

Here, T$_A$ is the dispersion of the spin fluctuation spectrum in wave-vector space. T$_A$, $M$, and $H$ are in K, $\mu_B$/Cr, and Oe units, respectively. Fig.~\ref{SCR}(b) shows the plot of  M$^4$ $vs.$ $H/M$, fitted by  Eq.~\ref{Eqn1}.  Such a linear behavior is observed in many itinerant ferromagnets such as MnSi~\cite{Chattopadhyay2009}, LaCo$_2$P$_2$~\cite{Imai2015}, SmCoAsO~\cite{Ohta2010}.  From the fitting of Eq.~\ref{Eqn1}, we obtained a slope of 9.4778$\times$10$^{-6}$ ($\mu_B$/Cr)$^5$/Oe. Using the slope and T$_C\approx170$ K, we derived T$_A\approx$ 868 K for $H\parallel c$.

Further, following the SCR theory, T$_C$ can be expressed by,

\begin{equation}
  T_C = (60c)^{-3/4}P_S^{3/2}T_A^{3/4}T_0^{1/4}
 \label{Eqn2}
 \end{equation}

Here, $c=$ 0.3353, P$_S$ is the spontaneous magnetization in $\mu_B$/Cr, T$_0$ is the energy width of the dynamical spin fluctuation spectrum in K. Using the value of T$_C$, P$_S$, and T$_A$, we obtain the characteristic temperature T$_0$ = 4502 K for our Cr$_{2}$Te$_{3}$ single crystal. According to the SCR theory of spin fluctuation, the ratio T$_C$/T$_0$ is an important parameter as it characterizes the degree of localization or itineracy of the spin moment. The magnetic materials are found to exhibit itinerant character for T$_C$/T$_0$ $\ll $ 1, while they show localized magnetism for T$_C$/T$_0\approx$ 1. In our case of Cr$_{2}$Te$_{3}$ single crystal, the ratio T$_C$/T$_0$ is estimated to be 0.04 which is much smaller than 1, indicating the itinerant nature of the ferromagnetism in Cr$_{2}$Te$_{3}$.

\subsection{Critical Behaviour Analysis}
Investigating the magnetic interactions across the paramagnetic to ferromagnetic transition involves the critical exponents analysis of magnetization isotherms [$M(H)$] around $T_C$. The $M(H)$ data measured for $H\parallel c$ between 150 and 190 K with an interval of 2 K are depicted in Fig.~\ref{3}(a). Fig.~\ref{3}(b) shows Arrott plots of $M^{2}$ $vs.$ $(H/M)$ for the Landau mean-field model. Arrott plots typically display a positive slope of parallel lines in the high-field region, indicative of a second-order magnetic transition~\cite{Arrott1957}. However, our findings show deviation from the linear behavior, particularly below T$_C$. This deviation suggests that the Landau mean-field theory ($\beta = 0.5, \gamma = 1$) is not applicable to Cr$_2$Te$_3$. Consequently, a modified Arrott plot (MAP) is necessary to determine the critical exponents by following the below equation,

\begin{equation}
    (H/M)^{1/\gamma} = a*(T - T_c)/T_c + b*M^{1/\beta}
    \label{Eq0}
\end{equation}
where $(H/M)$ is the inverse susceptibility, $M$ is the magnetization, $T_C$ is Curie temperature, $\beta$ and $\gamma$ are the critical exponents,  and $a$ and $b$ are constants.

Scaling hypothesis of the critical phenomena theory~\cite{Stanley1971} suggests that the second-order phase transitions near T$_C$ are governed by the critical exponents and magnetic equations of the state. The divergence of the correlation length around $T_C$, $\zeta = \zeta_0[(T - T_C)/T_C]^{-\nu}$, gives rise to universal scaling laws. This results in specific critical exponents $\beta$ for spontaneous magnetization $M_{sp}$ below T$_C$, $\gamma$ for the inverse susceptibility $\chi_0^{-1}$ above T$_C$, and $\delta$ for magnetization isotherms $M(H)$ at T$_C$ which are mathematically represented in the magnetic equations involving these critical exponents,

\begin{equation}
M_{sp}(T) = M_0(-\epsilon)^\beta \text{ for } \epsilon < 0, T < T_C,
\label{Eq1}
\end{equation}
\begin{equation}
\chi_0^{-1}(T) = (h_0/m_0)\epsilon^\gamma, \text{ for } \epsilon> 0, T > T_C
\label{Eq2}
\end{equation}
\begin{equation}
 M = DH^{1/\delta}, \text{ for } \epsilon = 0, T = T_C,
 \label{Eq3}
\end{equation}

 \begin{table*}[hbt!]
\caption{Comparison of critical exponents of Cr$_2$Te$_3$ with similar 2D magnetic materials and different theoretical models.}
\begin{tabular*}{0.9\linewidth}{c @{\extracolsep{\fill}} ccccccc}
 \hline
Composition & Technique & T$_{c}$ & $\beta$ & $\gamma$ & $\delta$ & Reference \\ [1.5ex]
 \hline\hline
Cr$_{2}$Te$_3$ & MAP & 170.21(1) & 0.360(2) & 1.221(3)& 4.392(6) &This work \\ [1.2ex]
 \hline
Cr$_{2}$Te$_3$ & KF method & 170.07(6) & 0.353(4) & 1.213(5)& 4.436(2) &This work  \\[1.2ex]
 \hline
Cr$_{2}$Te$_3$ & Critical isotherm & 170 &  & & 4.275 & This work \\[1.2ex]
 \hline
 Theory & 3D Ising  &  & 0.325 & 1.241 & 4.82 & \cite{Kaul1985}\\[1.2ex]
 \hline
 Theory & 3D XY  &  & 0.345 & 1.316 & 4.81 & \cite{Kaul1985}\\[1.2ex]
 \hline
 Theory & 3D Heisenberg  &  & 0.365 & 1.386 & 4.80 & \cite{Kaul1985}\\ [1.2ex]
 \hline
Theory & Landau mean-field  &  & 0.5 & 1 & 3 & \cite{Arrott1957} \\[1.2ex]
 \hline
 Theory & Tricritical mean-field  &  & 0.25 & 1 & 5 & \cite{Kim2002} \\[1.2ex]
 \hline
 Cr$_{5}$Te$_8$  & KF method & 230.6(3) &0.315(7) & 1.81(2) & 6.35 & \cite{Liu2017a}\\[1.2ex]
 \hline
 Cr$_3$Te$_4$  & KF method & 320.335(5) &0.365(5) & 1.212(9) & 4.580(4) & \cite{Wang2023}\\[1.2ex]
 \hline
 Cr$_4$Te$_5$  & KF method & 319.06(9) &0.387(9) & 1.2885(2) & 4.32 & \cite{Zhang2020}\\[1.2ex]
 \hline
 Cr$_5$Te$_6$  & KF method & 338.17(1) &0.405(1) & 1.200(1) & 3.962(9) & \cite{Zhang2022}\\[1.2ex]
 \hline
\end{tabular*}
\label{T1}
\end{table*}

where $\epsilon$ = (T – T$_C$)/T$_C$ is the reduced temperature, $M_0$, $h_0/m_0$, and $D$ are the critical amplitudes~\cite{Fisher1967}. Attempts were made to derive the critical exponents and T$_C$ through MAP technique by plotting M$^{1/\beta}$ $vs.$ (H/M)$^{1/\gamma}$ based on various theoretical models around T$_C$ such as the 3D-Heisenberg model, 3D-XY model, 3D-Ising model, and tricritical mean-field model~\cite{LeGuillou1980,Kaul1985,Kim2002} as shown in Figs.~\ref{3}(c), ~\ref{3}(d), ~\ref{3}(e), and ~\ref{3}(f), respectively. However, none of these models produced the anticipated parallel lines. To further confirm that none of these models can adequately describe the magnetic interactions of this system, in Fig.~\ref{slope} we plotted the normalized slopes (NS), $NS=S(T)/S(T_C)$,  as a function of temperature at higher magnetic fields derived from the data shown in Figs.~\ref{3}(b)-~\ref{3}(f) for various models.  Here, $S(T)$ is the slope defined by  $d(M^{1/\beta})/d((H/M)^{1/\gamma})$ at a given temperature $T$. For any adequate model, the normalized slopes should be equal to one~\cite{Liu2017a, Wang2023}. However, from Fig.~\ref{slope} it is evident that none of the models show NS=1 at temperatures other than $T_C$. Therefore, none of the standard models, Landau mean-field, 3D-Heisenberg, 3D-XY, 3D-Ising, and tricritical mean-field model,  can describe the magnetic interactions in Cr$_2$Te$_3$ precisely.

Thus, to accurately determine the critical exponents near T$_C$, an iterative modified Arrott plot (MAP) technique was employed on the data shown in Fig.~\ref{3}(a)~\cite{Arrott1967,Pramanik2009}. In this method, we start with trial values of $\beta$ and $\gamma$ substituted into Eq.~\ref{Eq0} to derive the MAP data as shown in Fig.~\ref{4}(a).  The linear extrapolation of the MAP data from the high field region provides $(M_{sp})^{1/\beta}$ and $(\chi_{0}^{-1})^{1/\gamma}$ as intercepts on the $y$ and $x$-axis of the $(M)^{1/\beta}$ vs. $(H/M)^{1/\gamma}$ plot [see Fig.~\ref{4}(a)]. As obtained $M_{sp}$ and $\chi_{0}^{-1}$ are plotted as a function of temperature as shown in Fig.~\ref{4}(c).  Next, $M_{sp}(T)$ and $\chi_0^{-1}(T)$  are fitted using Eqs.~\ref{Eq1} and ~\ref{Eq2} to determine new $\beta$ and $\gamma$ values. Using these updated $\beta$ and $\gamma$ exponents, a new MAP is created by plotting $(M)^{1/\beta}$ vs. $(H/M)^{1/\gamma}$. This procedure is iteratively repeated, involving fitting the modified Arrott plots, updating the polynomials (Eqs.~\ref{Eq1} and ~\ref{Eq2}), and recalculating the exponents, until the values of $\beta$ and $\gamma$ converge ($\beta_{n+1}\rightarrow\beta_n$ and $\gamma_{n+1}\rightarrow\gamma_n$, here $n$ is the iteration number). After this iterative exercise, Fig.~\ref{4}(a) shows parallel straight lines with $\beta=$ 0.367 and $\gamma=$ 1.263 in which the straight line corresponding to the $T_C$ passes through the origin. Further, from the converged MAP, the $M_{sp}(T)$ and $\chi_0^{-1}(T)$  are plotted in Fig.~\ref{4}(c) from which the fittings with Eqs.~\ref{Eq1} and ~\ref{Eq2} derive the critical exponents $\beta$ = 0.360(2) with T$_C$ = 170.21(1) K and $\gamma$ = 1.221(3) with T$_C$ = 170.16(4) K. These $T_C$ values closely match the T$_C$ of 170 K estimated from $M(T)$ data [see Fig.~\ref{2}(a)].


To verify the accuracy and reliability of the critical exponents determined by the modified Arrott plot method, we employed the Kouvel-Fisher (KF) method~\cite{Kouvel1964}, using the equations,

\begin{equation}
M_{sp}(T)[dM_{sp}(T)/dT]^{-1} = (T - T_C)/\beta,
\label{Eq4}
\end{equation}

\begin{equation}
 \chi_0^{-1}(T)[d \chi_0^{-1}(T)/dT]^{-1} = (T - T_C)/\gamma,
 \label{Eq5}
\end{equation}

Thus, Fig.~\ref{4}(d) exhibits KF plots fitted using the Eqs.~\ref{Eq4} and \ref{Eq5} both below and above $T_C$. In this way, we obatined the critical exponents $\beta$ = 0.353(4) with $T_C$ = 170.16(2) K and $\gamma$ = 1.213(5) with $T_C$ = 170.07(6) K, respectively. Notably, these values are very close to the critical exponents obtained from the MAP method, affirming the reliability, self-consistency,  and intrinsic nature of the critical exponents. Furthermore, using the Widom scaling relation, $\delta$ = 1 + $(\gamma$/$\beta)$ \cite{Widom1964}, we estimate $\delta$ = 4.392(6) and $\delta$ = 4.436(2) from the MAP and KF plots, respectively. Crucially, these $\delta$ values, obtained from different methods converge to $\delta$ = 4.275(10) derived through the critical isotherm (CI) method using Eq.~\ref{Eq3} [see Fig.~\ref{4}(b)]. The experimentally derived critical exponents are summarized in Table~\ref{T1} along with critical exponents of different theoretical models and other Cr$_x$Te$_y$ systems.

Further, in the critical asymptotic region of magnetic materials, the scaling equation of magnetic state is expressed by~\cite{Stanley1971},
 \begin{equation}
        M (H, \epsilon) = \epsilon^\beta f_\pm (H/\epsilon^{\beta + \gamma})
        \label{Eq6}
 \end{equation}
 This equation can also be expressed as,
\begin{equation}
    m = f_\pm (h)
     \label{Eq7}
\end{equation}

where $m = \epsilon^{-\beta} M (H, \epsilon) $ denotes the normalized magnetization and $h =  H\epsilon^{-(\beta + \gamma)}$ represents the normalized field. When the values of $\beta$ and $\gamma$  obtained from both the MAP and KF methods are accurately chosen, the scaled magnetization $m$ and $h$ are expected to mainly fall into to two distinct branches of universal curves outlined by the Eqs.~\ref{Eq6} and~\ref{Eq7}. These branches correspond to temperatures below and above T$_C$ as evident in Figs.~\ref{4}(e) and \ref{4}(f).

 \begin{table}[b]
\caption{Obtained critical parameters from renormalization group theory}
\begin{tabular*}{\linewidth}{c @{\extracolsep{\fill}} ccccc}
 \hline
\{d : n\} & $\gamma$ & $\beta$ & $\sigma$ & $\delta$ \\ [1.5ex]
 \hline\hline
\{3 : 1\} & 1.213 & 0.362 & 1.888 & 4.395\\[1.2ex]
 \hline
\{3 : 2\} & 1.213 & 0.395 & 1.827 & 4.116\\[1.2ex]
 \hline
 \{3 : 3\} & 1.213 & 0.415 & 1.791 & 3.964 \\[1.2ex]
 \hline
\end{tabular*}
\label{T2}
\end{table}

The compiled critical exponents in Table~\ref{T1}, compared with theoretical predictions across various models, reveal complex magnetic interactions across the Curie temperature in Cr$_2$Te$_3$. Specifically, the value of $\beta$ ($T<T_C$) obtained from the MAP technique is very close to the 3D-Heisenberg type interactions, while the value of $\gamma$ ($T>T_C$) is very close to the 3D-Ising type interactions.  However, the value of $\delta$ obtained at $T_C$ does not fall into any of the existing standard models of 3D-Heisenberg, 3D-XY, or 3D-Ising (See Table~\ref{T1}). The deviation of the magnetic interactions may stem from various factors such as long-range Ruderman-Kittel-Kasuya-Yosida (RKKY)  interactions~\cite{Kaul1985}, extended interactions beyond nearest neighbors~\cite{Kaul1981}, dipole-dipole interaction~\cite{Tateiwa2014}, or strong magnetocrystalline anisotropy (MAE)~\cite{Yang2021,Liu2021,Yang2022}. Thus, it is necessary to understand the range of magnetic interactions and spin dimensionality of Cr$_2$Te$_3$. For a homogeneous magnet, the magnetic phase transition's universality class depends on the exchange interaction $J(r)$ and following the renormalization group (RG) theory the interactions decay with distance as $J (r) = r^{-(d + \sigma)}$.   Here, $d$ is the spacial dimensionality and $\sigma$ is a positive constant. The relation between the critical exponent $\gamma$ and $\sigma$ can be expressed by~\cite{Fisher1967,Fischer2002},

 \begin{equation}
 \begin{split}
  \gamma & = 1 + \frac{4}{d}\frac{(n + 2)}{(n + 8)}\Delta\sigma +\frac{8(n - 4)(n + 2)}{(n + 8)^2d^2}\\
         &  * [\frac{2(7n + 20)G(d/2)}{(n + 8)(n - 4)}+1 ]\Delta\sigma^2
     \end{split}
     \label{Eq8}
 \end{equation}

 where $\Delta\sigma = \sigma - d/2$, $ G(d/2) = 3 -0.25*(d/2)^2$. $n$ and $d$ are the spin and spatial dimensionality of the system \cite{Fischer2002}. For $\sigma>$ 2, the short-range Heisenberg model is valid for describing a 3D-isotropic magnet with magnetic interactions $J(r)$ decaying faster than r$^{-5}$. Conversely, for $\sigma\leq$ 3/2, the system follows the long-range mean-field model with magnetic interactions $J(r)$ decaying slower than r$^{-4.5}$~\cite{Fisher1972}. For $3/2\le \sigma \le 2$ the system falls into a distinct category characterized by the critical exponents that vary according to the specific value of $\sigma$. According to Eq.~\ref{Eq8}, it is found that \{d : n\} = \{3 : 1\} and $\gamma$ = 1.213 give the critical exponents $\beta$ = 0.362, $\sigma$ = 1.89, and $\delta$ = 4.395 which are close to our experimental observations. The other exponents $(\nu, \alpha, \gamma)$ can be derived from the following scaling equations,  $\nu=\gamma/\sigma$, $\alpha=2-\nu*d$, $\beta=(2-\alpha-\gamma)/2$, where $\nu$ and $\alpha$ are the exponents of correlation length. This is done for various sets of \{d : n\} values, as shown in Table \ref{T2}. Thus, the exchange interactions in Cr$_2$Te$_3$ decay as $J (r) = r^{-(d + \sigma)}= r^{-4.89}$. The spin dimensionality $n =$ 1 and the spacial dimensionality $d =$ 3 as derived from the RG theory suggests 3D-Ising type magnetic interactions, stabilized by the robust uniaxial anisotropy present in the system. This behavior mirrors the observations in other Cr$_x$Te$_y$ systems~\cite{Liu2017a,Zhang2018}. Therefore, in Cr$_2$Te$_3$ the magnetism primarily arises from the intralayer coupling~\cite{Youn2007}, as evidenced by the 3.9254 Å bond length between adjacent Cr atoms within the plane, indicative of a relatively weak direct exchange interaction~\cite{Jiang2022}. Moreover, the Cr2-Te1-Cr3 bond angle within the layer, measuring approximately 91.213$^\circ$, indicates super-exchange interactions that favor ferromagnetism~\cite{Goodenough1955,Kanamori1959}. Further the strong hybridization between Cr $3d$ and Te $5p$ orbitals potentially establishes an indirect exchange between the Cr moments mediated by neighboring Te atoms~\cite{Dijkstra_1989}.

\begin{figure}[t]
    \centering
    \includegraphics[width=\linewidth]{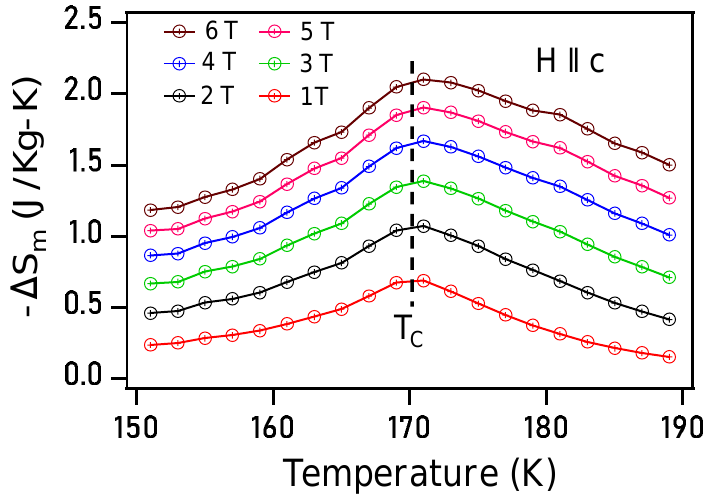}
    \caption{Temperature dependent  magnetic entropy change -$\Delta S_m$ at different applied magnetic fields.}
    \label{5}
\end{figure}

 Finally, before concluding this section, we would like to mention that during the course of our manuscript preparation a preprint discussing the magnetocrystalline anisotropy and magnetic interactions of Cr$_{1.77}$Te$_3$ (claimed as Cr$_2$Te$_3$) single crystals has appeared~\cite{Goswami2023}. Though, our results of out-of-plane magnetization matches well with the Ref.~\cite{Goswami2023}, the in-plane magnetization is slightly different. Moreover, based on the magnetocaloric effect (MCE) analysis,  Ref.~\cite{Goswami2023} suggests a 2D-Ising type magnetic interactions in their studied composition of Cr$_{1.77}$Te$_3$ despite the experimentally obtained $\beta=0.2435$ and $\gamma=1.6509$ are quite far from the theoretically  predicted values ($\beta=0.125$ and $\gamma=1.75$) for a 2D-Ising type magnet. Whereas, our studies demonstrate a complex magnetic interactions falling in between the 3D-Ising and 3D-Heisenberg models. We suggest that the difference in the type of magnetic interactions between our study and Ref.~\cite{Goswami2023} is possibly related to the differing Cr concentrations.

\subsection{ Magnetocaloric Effect}

The magnetocaloric effect that is inherent to ferromagnetic system, induces heating or cooling when the system is adiabatically subjected to the external magnetic fields~\cite{Pecharsky1999}. This phenomenon results in a magnetic entropy change -$\Delta S_m (T, H)$ in the presence of magnetic fields, expressed by the formula,

\begin{equation}
\Delta S_m (T, H) =  \int_{o}^{H} (\frac{\partial S}{\partial H})_T dH = \int_{o}^{H} (\frac{\partial M}{\partial T})_H dH
 \label{Eq9}
 \end{equation}

According to the Maxwell’s relation, ($\frac{\partial S}{\partial H})_T$ = ($\frac{\partial M}{\partial T})_H$. For the magnetization measured within small discrete fields and temperature intervals, $\Delta S_m (T, H)$ can be expressed as
\begin{equation}
\Delta S_m (T, H) = \frac{\int_{0}^{H} M(T_{i+1}, H) dH - \int_{0}^{H} M(T_{i}, H) dH }{T_{i+1}-T_i}
 \label{Eq10}
\end{equation}

Fig.~\ref{5} displays the temperature dependence of derived -$\Delta S_m (T, H)$ for $H\parallel c$ at different magnetic field strength.  These curves showcase a peak change in entropy around T$_C$, presenting a broad peak pattern. At 6 T of applied field -$\Delta S_m (T, H)$ reaches maximum of approximately 2.08 J kg$^{-1}$ K$^{-1}$ for $H\parallel c$. The observed -$\Delta S_m (T, H)$ values at 6 T are comparable to those obtained from other 2D magnetic materials such as Cr$_2$Ge$_2$Te$_6$ (2.64 J kg$^{-1}$ K$^{-1}$) \cite{Liu2019a} and Cr$_5$Te$_8$ (2.38 J kg$^{-1}$ K$^{-1}$) \cite{Liu2019}, larger than Fe$_{3-x}$GeTe$_2$ (1.14 J kg$^{-1}$ K$^{-1}$) \cite{Liu2019b} and CrI$_3$ (1.56 J kg$^{-1}$ K$^{-1}$) \cite{Liu2018}, and smaller than CrB$_3$ (7.2 J kg$^{-1}$ K$^{-1}$) \cite{Yu2019} and Cr$_2$Si$_2$Te$_6$ (5.05 J kg$^{-1}$ K$^{-1}$) \cite{Liu2019a}.

\section{Summary}

 We have successfully grown high-quality single crystals of Cr$_2$Te$_3$ using the chemical vapor transport method.  Our investigation on the critical behavior of Cr$_2$Te$_3$ across the ferromagnetic to paramagnetic (FM-PM) transition temperature of 170 K, reveals crucial insights into the magnetic exchange interactions. Specifically, the Kouvel-Fisher (KF) method derives the critical exponents, $\beta$ = 0.353(4) and $\gamma$=1.213(5),  suggesting complex magnetic interactions falling in between the 3D-Ising and the 3D-Heisenberg models.  Further, the renormalization group theory analysis indicates a 3D-Ising-type magnetic interactions decaying with the distance as $J(r) = r^{-4.89}$. Importantly, the uniaxial magnetocrystalline anisotropy (MAE) of $K_u=2065$ kJ/m$^3$ is the highest ever known experimental value in Cr$_x$Te$_y$ type systems. We suggest that the 3D-Ising type magnetic interactions in Cr$_2$Te$_3$ are stabilized by the extremely large uniaxial MAE. Further, the DFT calculations predict large MAE of $K_u=8230$ kJ/m$^3$, leading to a ground state noncollinear ferromagnetic structure with dominant out-of-plane ferromagnetic and dominant in-plane antiferromagnetic Cr-spin arrangements. The self-consistent renormalization theory (SCR) suggests Cr$_2$Te$_3$ to be an out-of-plane itinerant ferromagnet. Investigating the magnetic entropy change, -$\Delta S_{M}^{max}$, as a function of temperature provided insights on the magnetocalaric effect in this system. These findings collectively lay a robust groundwork for advancing the magnetocaloric and spintronic technologies.

\section{Acknowledgements}

S.T. thank the Science and Engineering Research Board (SERB), Department of Science and Technology (DST), India for the financial support (Grant No. SRG/2020/000393). This research has made use of the Technical Research Centre (TRC) Instrument Facilities of S. N. Bose National Centre for Basic Sciences, established under the TRC project of the Department of Science and Technology, Govt. of India.

\bibliographystyle{model1-num-names}
\bibliography{Cr2Te3.bib}

\end{document}